# Monitoring Networked Applications With Incremental Quantile Estimation[1]

**John M. Chambers, David A. James, Diane Lambert and Scott Vander Wiel**

*Abstract.* Networked applications have software components that reside on different computers. Email, for example, has database, processing, and user interface components that can be distributed across a network and shared by users in different locations or work groups. End-to-end performance and reliability metrics describe the software quality experienced by these groups of users, taking into account all the software components in the pipeline. Each user produces only some of the data needed to understand the quality of the application for the group, so group performance metrics are obtained by combining summary statistics that each end computer periodically (and automatically) sends to a central server. The group quality metrics usually focus on medians and tail quantiles rather than on averages. Distributed quantile estimation is challenging, though, especially when passing large amounts of data around the network solely to compute quality metrics is undesirable. This paper describes an Incremental Quantile (IQ) estimation method that is designed for performance monitoring at arbitrary levels of network aggregation and time resolution when only a limited amount of data can be transferred. Applications to both real and simulated data are provided.

*John M. Chambers retired in 2005 as Member of the Technical Staff, Communications and Statistics Research, Bell Labs, Lucent Technologies, Murray Hill, New Jersey 07974, USA e-mail: jmc@r-project.org. David A. James is a Member of the Technical Staff, Communications and Statistics Research, Bell Labs, Lucent Technologies, Murray Hill, New Jersey 07974, USA e-mail: dj@bell-labs.com. Diane Lambert is a Research Scientist, Google, New York, New York 10018, USA e-mail: dlambert@google.com. Scott Vander Wiel is a Technical Staff Member, Statistical Sciences Group, MS F600, Los Alamos National Laboratory, Los Alamos, New Mexico 87545, USA e-mail: scottv@lanl.gov.*

## 1. MONITORING NETWORKED APPLICATIONS

A stand-alone software application like a text processor resides entirely on one computer and is accessed only by the people who use that computer. The components and users of a networked software application like email, though, span multiple computers. The database that stores current email messages may reside on one (or more) computers, the database of previously read messages may reside on another computer, the mail processing software may reside on yet another computer, and the user in-terface that allows email to be read and sent easily may reside on many personal computers. That is, the components of the networked software, the users of the software, and the requests and actions by users are all distributed over the network.

Networked services can fail in many ways, and the failures are often localized to a set of nodes that share a small fraction of the network infrastructure. Email transactions for only a subset of users may be delayed by server problems that disrupt a region of a network, or database accesses may be slow because of heavy seasonal tasks that are performed by only some of the workers. Consequently, system administrators need to assess availability, reliability and performance with the structure of the network in mind, without specifying in advance which pieces of the network or which work groups to monitor together.

Monitoring the health of networked applications is challenging. First, the desktop computers or end user nodes that access the application may have only limited resources to allocate to processing metrics. At best, each end user may be able to compute limited summaries of its performance. Moving all the performance data concerning all transactions from all end users to a dedicated server does not circumvent the problem of weak end nodes because transferring large amounts of data can place too high a load on the network. Thus, both the data and computational resources needed to compute quality metrics for networked software applications need to be distributed over the network. Finally, there are statistical challenges too. For example, users in the same building may have dissimilar tasks, so the aggregated performance data from that location look like a sample from a mixture with multiple modes and long tails rather than like a sample from a simple parametric model.

This paper describes an approach to monitoring networked applications that we developed in response to the needs of a business unit of Lucent Technologies. To accommodate a wide range of statistical distributions, monitoring is based on tracking medians and upper quantiles rather than averages and higher-order moments. The nature of the specific problem, the constraints on computing that have to be addressed, and a high-level view of the approach we took are described in Section 2; related approaches are discussed in Section 3. Our design has two parts: a lightweight sequential method that summarizes the performance data that are collected at each user's computer (Section 4) and a slight variant of the sequential method that further aggregates the user summaries over arbitrary subsets of the network and time (Section 5). (Using nearly the same algorithm at the end-user and server levels was one of the constraints specified by the engineers of our application.) Enhancements to achieve better accuracy are discussed in Section 6. Performance of the user-level algorithm is evaluated on simulated data (Section 7). Performance of the server algorithm that computes group-level metrics is evaluated on transaction time data collected from a group of corporate users and simulated work-group data (Section 8). Some ideas for generalizing the methods are given in Section 9.

## 2. MONITORING NETWORKED SOFTWARE

Networked software provides applications such as email, database access, and voice and conferencing services to an enterprise. In a typical configuration, portions of the software live on *servers* and employees of the enterprise access it using *clients* that live on their desktop computers. *Monitoring agents* are special clients that observe the performance details for each attempted and completed software transaction: round trip time, server response time, bandwidth used, completion status, packet loss, total transaction time, and so on. It is these performance data that describe the software quality that the user has experienced, and the data for a group of users describe the software quality delivered to the group. The monitoring agents summarize the data and periodically send the summaries to a central server that is responsible for monitoring the reliability and performance of the application across the network. Figure 1 illustrates the high-level flow of data and summary records in the monitoring application. In these applications, reliability problems are failures of the network, servers and applications to deliver adequate performance to the end users. Problems may not be exhibited through complete failure of the infrastructure, but rather through soft metrics such as overly long response times on high volume transactions.

To save space on the end user's computer, the monitoring agents summarize the performance data with a fixed-length record, one record for each transaction type, that is updated with new performance data whenever the networked application is used. Often the record is too small to hold all the raw data, and in this case it must hold summaries of the data rather than the full set of data values. Periodically, say at the end of every hour, the summary record is sent to a server. The server then aggregates the summary records across locations, work groups, business units and longer periods of time as required by system administrators investigating reliability and performance issues. Server records are also fixed-length.

Figure 2 shows a histogram of times to complete email transactions with SMTP or POP3 servers aggregated over 15 employees in a one-month period. The shortest transaction time is 1 ms, while the longest is $2.33 \times 10^5$ ms or 233 seconds. No standard transformation of these data induces normality or even symmetry. Moreover, as would be expected when aggregating over agents and times, the histogram for the work group is multimodal.

Summarizing such data quickly and reliably while preserving as much information as possible about the entire distribution is especially challenging because the transaction times are obtained sequentially across a group of end users, there is not enough memory to store all the data for many metrics on many transaction types before they are analyzed, and the data cannot be reduced to a small set of sufficient statistics by appealing to a parametric family of distributions. Simple statistical summaries such as the mean and variance are statistically inadequate (unfortunately so, since they are inexpensive to compute). Under these circumstances, we prefer to summarize the distribution in terms of its median and tail quantiles.

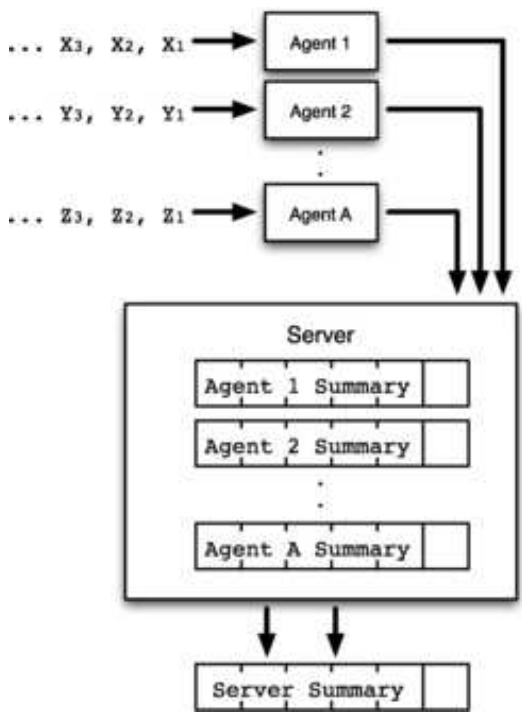

Fig. 1. *Data flow for monitoring networked software.*

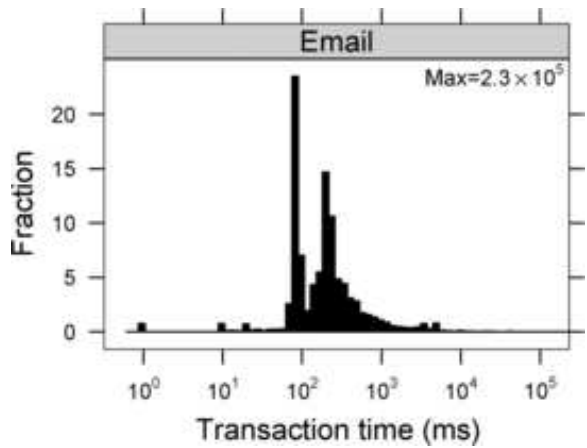

Fig. 2. *Times to complete* 41,928 *email transactions over a one-month period.*

## 3. INCREMENTAL QUANTILES

In statistical notation, agent $a$ (the agent monitoring your computer, say) sees a multivariate *data stream*

$$\mathbf{X}_a = \{X_{ast}, s = 1, \ldots, S, t = 1, 2, \ldots\},$$

where $X_{ast}$ is the value of the $s$th metric (response time, e.g.) on the $t$th transaction (email access, e.g.) seen by agent $a$.

Users of the software application are typically organized in multiple hierarchies according to geographic location and business unit. The interesting subsets of agents correspond to these hierarchies or to groups defined by common network infrastructure. Time adds another dimension, and the interesting periods may be five-minute periods, hours, days or months depending on the purpose of the analysis. Often, the agent hierarchy and time resolution are chosen dynamically as an analyst explores the data. But whatever the choices, the analyst is to be provided quantiles for the aggregated data $\{X_{ast} : a \in \mathcal{A}, s \in \mathcal{S}, t \in \mathcal{T}\}$ where $\mathcal{A}$, $\mathcal{S}$ and $\mathcal{T}$ are subsets of agents, metrics and time, respectively. Quantile estimates for the aggregate data are produced from records that are periodically provided by agents. Each of the agent records in turn contains a set of quantile estimates that were produced by that agent using the same kind of sequential updating algorithm that the server uses.

Sequential quantile estimation, which is called incremental quantile estimation in the computer science literature, is not a new topic. Robbins and Monro (1951) introduced the idea of stochastic approximation for quantile estimation, for example. Munro and Paterson (1980) then used it for sorting and selection with limited memory, Tierney (1983) used it for monitoring computer simulations, and Chen, Lambert and Pinheiro (2000) used it for monitoring nonstationary user profiles. Stochastic approximation is best suited for continuous data because it requires an estimate of the density near the quantile. The data in our application, such as packet sizes, are often discrete and can often have preferred values and spikes, so any continuity assumption is suspect. Liechty, Lin and McDermott (2003) proposed an algorithm to estimate a single quantile by maintaining a buffer of data values that is intended to bracket the desired quantile. Their algorithm works well for simulated data, but it tracks only a single quantile. McDermott, Babu, Liechty and Lin (2003) extended the algorithm to track a prespecified set of quantiles. The Incremental Quantile (IQ) method represents a different emphasis, on estimating distribution functions as a whole and combining those estimates for a general data-analytic tool. Future numeric comparisons with alternative algorithms such as those referenced above may lead to improved estimates within this general approach.

Computer scientists have considered sequential quantile estimation without density estimates, but with the twist that reported quantiles must be observed data values. See Manku, Rajagopalan and Lindsay (1998) and Greenwald and Khanna (2001, 2004). Simply stated, these methods attempt to keep "typical" values, so that the goal is perhaps more akin to sorting the data than to estimating an underlying distribution. Our application does not have the constraint that quantile estimates must be observed data values. The advantage of the computer science methods is that they guarantee precision to within a prespecified error on the probability level of the quantile estimate. Such guarantees can be useful, but much less so when interest is in tail quantiles. For example, it may be adequate to estimate the median to within the interval defined by the 0.49 and 0.51 empirical quantiles, but a fixed ±0.01 error on the probability level is nearly useless for estimating the 0.999 quantile. In our application, interest centers on the accuracy of the estimated quantile *value* itself rather than its *probability level*.

Three simple principles underlie our approach to sequentially estimating and aggregating quantiles:

1. Empirical distributions are appropriate for all sorts of numerical data.
2. Averaging cumulative distribution functions (CDFs) is easy.
3. Converting a CDF to a set of quantiles and vice versa is straightforward.

To aggregate sets of quantiles provided by many agents, we collect a batch of agent records until a fixed number has been reached, and then convert the quantiles on the records to empirical CDFs and the quantile record at the server to another CDF. Then we average the CDFs with appropriate weights and compute quantiles of the average CDF to complete one round of the aggregation algorithm. Of course, the way that a set of quantiles is converted to a CDF may affect the quality of the final estimates, as does the choice of the probability levels for the quantiles in each set. This procedure is simple, but it seems not to have been used previously. Details and performance comparisons are provided in the remainder of this paper.

## 4. IQ AGENT ALGORITHM

### 4.1 Requirements for Aggregation Algorithms

The monitoring architecture requires two types of algorithms, one for the agent and one for the server. The agent algorithm should require only one continuous pass through the data stream and should be lightweight in both memory and CPU usage because many copies of the algorithm (one for each transaction type for each networked application and monitored quantity) will run in the background on the desktops of corporate users. Hourly records produced by the algorithm should be fixed-length to simplify the design and small to reduce the burden of transmitting them to the server for further aggregation.

Figure 3 depicts the major steps in the IQ agent algorithm. A data buffer $\mathbf{D}$ at the agent holds the most recent observations from a stream $\{X_1, X_2, \ldots\}$. A quantile buffer $\mathbf{Q}$ corresponding to probability values $\mathbf{P}_Q = (p_1, \ldots, p_M)$ holds the quantiles $\mathbf{Q} = (Q_1, \ldots, Q_M)$ estimated from the data that have already been processed. When $\mathbf{D}$ fills with data, it is first used to update **Q** and then it is cleared in order to accumulate the next batch of data from the stream. When a report is required, a predetermined subset of **Q** is provided to the server as a summary of the entire stream processed by the agent. Notice that more quantiles may be tracked in the **Q**-buffer than are reported in the agent summary to improve the accuracy of the agent record.

At the server, a second algorithm summarizes agent records by estimating quantiles of the mixed distribution of their combined data. Like the agent algorithm, the server algorithm should be lightweight and operate in one pass through a set of agent records. Ideally, the server algorithm should create records of the same form as agent records to keep the design simple and to provide a uniform method for aggregating in stages up the levels of a hierarchy.

Details of the agent algorithm are provided in the remainder of this section. The server algorithm is discussed in Section 5.

### 4.2 Updating the Q-Buffer

Suppose that $T$ data values have been processed with the IQ algorithm so that **Q** holds estimated quantiles of the set $\{X_1, \ldots, X_T\}$. Then the data buffer **D** is filled with the next $N$ values, $\{X_{T+1}, \ldots, X_{T+N}\}$. When full or at prespecified times, **D** is converted to an empirical CDF $F_D(x)$, **Q** is converted to a CDF $F_Q(x)$, and a weighted average of the two CDFs is computed. Quantiles of the average CDF are used to update **Q**.

Linearly interpolating $F_Q$ models the data as uniformly distributed between adjacent quantiles in **Q**, which is reasonable if no other information is available and the tails of the data are not overly long. If a variable such as round-trip time or transaction time has a long right tail, then accuracy is improved by applying the algorithm to logged data or by using nonlinear interpolation as described in Section 6.

The updating algorithm has four basic steps, illustrated in Figure 4 and detailed as follows.

For each $x \in \mathbf{Q} \cup \mathbf{D}$:

1. Compute the CDF of **Q** (Figure 4, left panel) as

$$(1) \quad F_Q(x) = \begin{cases} 0, & \text{if } x < Q_1, \\ 1, & \text{if } x \geq Q_M, \\ \texttt{interp}(x, Q_m, Q_{m+1}, p_m^*, p_{m+1}^*), \\ \quad \text{if } Q_m \leq x < Q_{m+1}, \\ \quad m = 1, \ldots, M-1, \end{cases}$$

where interp interpolates the given points as

$$\texttt{interp}(x, x_0, x_1, p_0, p_1)$$
$$= p_0 + (p_1 - p_0) \frac{x - x_0}{x_1 - x_0}$$

(see Section 6 for nonlinear interpolation) and

$$p_m^* = \text{median}(p_m, 0.5/T, 1 - 0.5/T),$$

which is $p_m$ trimmed to the interval $[0.5/T, 1 - 0.5/T]$. Trimming imposes jumps in the CDF at the minimum ($Q_1$) and maximum ($Q_M$) data values, so the minimum and maximum over all data values processed so far are kept in **Q**. This means that half of the $1/T$ mass associated with an extreme value (minimum or maximum) is allocated to an interval strictly less extreme than the observed value, and the other half of the $1/T$ mass is allocated to the extreme value itself. It may be reasonable to replace the jump with a smooth extrapolation, but then some extreme quantiles would extend beyond the range of the observed data, which we choose to avoid.

2. Compute the empirical CDF of **D** (Figure 4, center panel) and its left-continuous value as

$$(2) \quad \begin{aligned} F_D^+(x) &= \frac{|\mathbf{D} \leq x|}{|\mathbf{D}|}, \\ F_D^-(x) &= \frac{|\mathbf{D} < x|}{|\mathbf{D}|}, \end{aligned}$$

where $|\cdot|$ indicates the number of elements in the indicated set.

3. Compute the weighted average CDF (Figure 4, right panel) and its left-continuous value as

$$F^\pm(x) = \frac{T \cdot F_Q(x) + N \cdot F_D^\pm(x)}{T + N}.$$

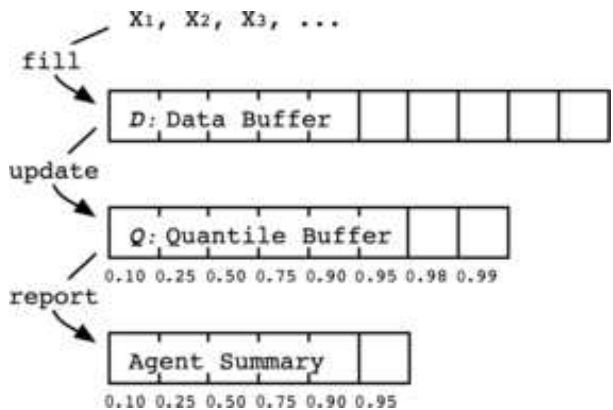

Fig. 3. *Major steps in the IQ agent algorithm.*

For each $p_m \in \mathbf{P}_Q$:

4. Compute the updated quantile, $Q_m$ (Figure 4, arrows in right panel), by inverting the weighted average CDF as follows. Find bracketing values

$$x^+ = \min_{F^+ \geq p_m} \{\mathbf{Q} \cup \mathbf{D}\},$$
$$x^- = \max_{F^- \leq p_m} \{\mathbf{Q} \cup \mathbf{D}\}$$

and set

$$(3) \quad Q_m = \begin{cases} x^-, & \text{if } x^+ = x^-, \\ \rho x^- + (1-\rho)x^+, & \text{otherwise,} \end{cases}$$

where $\rho = [F^+(x^+) - p_m]/[F^+(x^+) - F^-(x^-)]$ for linear interpolation. The nonlinear case is discussed in Section 6.

Finally, refill $\mathbf{Q}$ with the updated quantiles and clear $\mathbf{D}$ in order to resume accumulating new data from the stream.

The quality of IQ quantile estimates depends on the quality of the estimate of the CDF $F$ (i.e., $F^\pm$) from which they are computed, which in turn depends on the buffer sizes and probability levels $\mathbf{P}_Q$. In particular, the assumed $F$ is linear between distinct adjacent quantiles (or linear on a transformed scale), and this may be a better assumption over small intervals than over long intervals. Thus, keeping more quantiles in $\mathbf{Q}$ is desirable, even if only a few quantiles can be reported ultimately.

When all the data have been processed, an *agent record* can be formed to summarize the results. The agent record is $(T_a, \mathbf{R}_a)$ where $T_a$ is the total number of observations processed and $\mathbf{R}_a$ is typically a fixed subset of the quantiles in $\mathbf{Q}$, including the minimum and maximum values. However, if $T_a$ is smaller than the record size, then all the raw data values are inserted into $\mathbf{R}_a$.

### 4.3 An Example of IQ Updating

As an example, consider the transaction time data shown in Figure 2. Empirical quantiles (EQ) were computed in the standard way by sorting all the test data, and IQ quantiles were computed using buffer sizes $|\mathbf{D}| = |\mathbf{Q}| = 100$ and linear interpolation on the logged data. However, even the logged data remained long-tailed. The probabilities in $\mathbf{P}_Q$ were 0 and 1 (corresponding to the minimum and maximum data values) and 98 probabilities uniformly spaced from 0.0025 to 0.9975 on the $\log(p/(1-p))$ scale, so that more quantiles are devoted to tail probabilities.

Table 1 shows the IQ and EQ estimates, their differences, and approximate EQ standard errors computed by plugging a local density estimate into the asymptotic standard error formula. The IQ estimates reproduce the EQ values well with differences never more than two standard errors of the empirical quantiles.

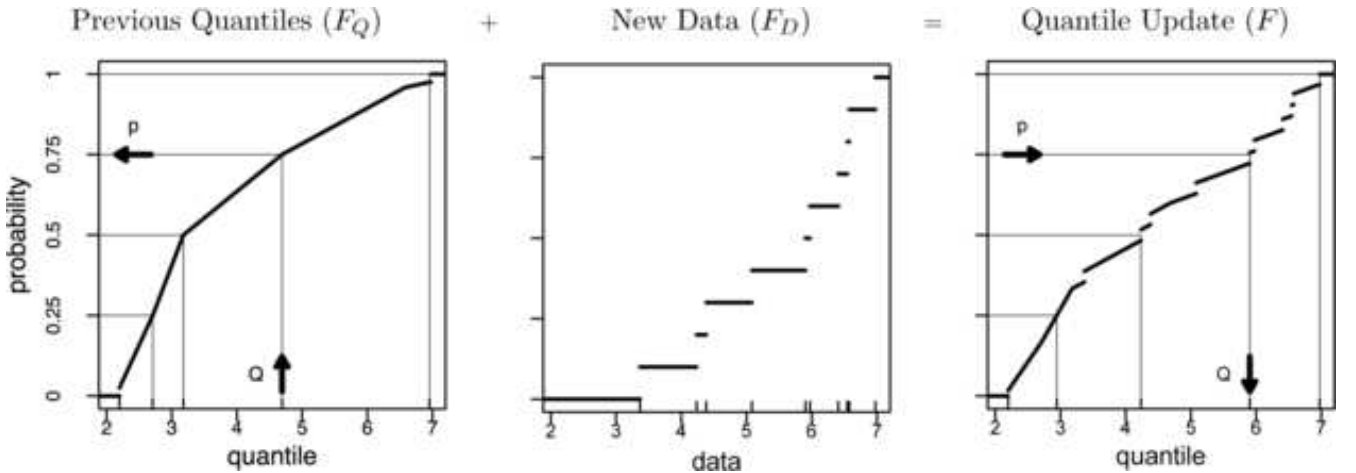

FIG. 4. *Quantile updating with $\mathbf{Q}$ of size 5 with probabilities $\mathbf{P}_Q = (0, 0.25, 0.5, 0.75, 1)$ and $\mathbf{D}$ of size 10. $\mathbf{Q}$ has been updated twice, so $T = 20$. The left plot shows $F_Q$ before updating where vertical segments indicate the stored quantiles. The middle plot shows the ten data values in $\mathbf{D}$ as ticks on the horizontal axis and the empirical CDF $F_D$. The right plot shows the updated $F$ (a weighted average of $F_Q$ and $F_D$). The updated quantiles for $\mathbf{Q}$ are shown as ticks along the horizontal axis.*

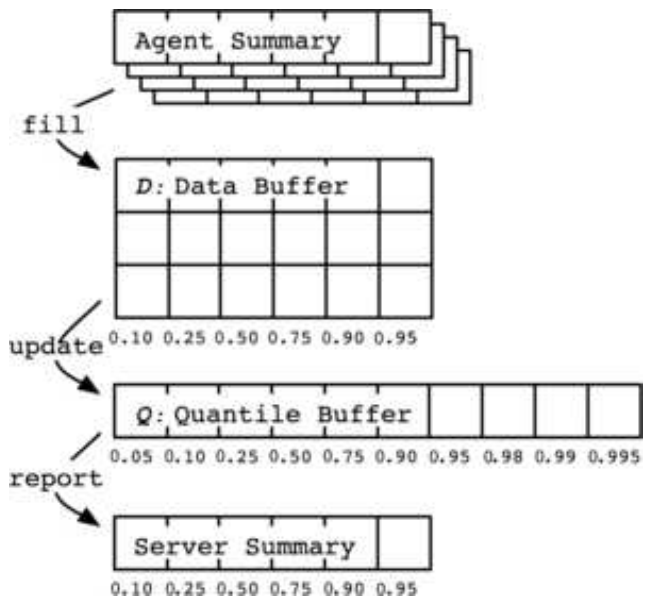

Fig. 5. *Major steps in the IQ server algorithm.*

## 5. IQ SERVER ALGORITHM

The next task is to merge sets of agent quantiles to estimate performance for a set of users, or to merge server quantiles to obtain estimates for combined work groups or longer periods of time, for example. To be specific, this section describes merging of agent records, but the ideas also apply to higher levels of aggregation. Figure 5 illustrates the major steps: agent summary records are placed into a data buffer $\mathbf{D}$; when $\mathbf{D}$ is full it is used to update a quantile buffer $\mathbf{Q}$; once all records have been processed, a subset of $\mathbf{Q}$ is selected to form a summary record of the aggregation.

As in the agent algorithm, $\mathbf{Q}$ holds the approximate quantiles $\mathbf{Q} = (Q_1, \ldots, Q_M)$ with corresponding probability levels $\mathbf{P}_Q$. These quantiles are a summary of all agent records that have been processed so far. When $\mathbf{Q}$ is updated, two ancillary quantities are also updated—$N_A$, the total number of agent records that have been processed and $T$, the total number of data values represented by the $N_A$ agents.

$\mathbf{D}$ holds the next set of agent records to be included in the aggregation, some of which contain quantiles and some of which may contain raw data values. The combined set of raw data values over all records in $\mathbf{D}$ is denoted by $\mathbf{X} = \{X_1, \ldots, X_N\}$. A quantile record from agent $a$ is denoted $(T_a, \mathbf{R}_a)$, where $T_a$ is the number of values represented and $\mathbf{R}_a = (R_{a,1} \leq \cdots \leq R_{a,I})$ is a vector of $I$ quantiles with probability levels $\mathbf{P}_R$, including both 0 and 1.

Updating $\mathbf{Q}$ at the server is similar to updating $\mathbf{Q}$ at the agent. Both $\mathbf{D}$ and $\mathbf{Q}$ are converted to CDFs, the CDFs are averaged, and then the average is inverted to update $\mathbf{Q}$.

For each $x \in \mathbf{Q} \cup \mathbf{D}$:

1. Compute $F_Q(x)$ using (1).
2. Compute the CDF, $F_a(x)$, of each set of agent quantiles using (1) with $\mathbf{R}_a$ and $\mathbf{P}_R$ in place of $\mathbf{Q}$ and $\mathbf{P}_Q$, respectively.
3. Compute the empirical CDF, $F_D^+(x)$, of the data values $\mathbf{X} \subset \mathbf{D}$ and its left-continuous value, $F_D^-(x)$, using (2) with $\mathbf{X}$ in place of $\mathbf{D}$.
4. Compute the weighted average CDF and its left-continuous value as

$$F^{\pm}(x) = \frac{T F_Q(x) + N F_D^{\pm}(x) + \sum_a T_a F_a(x)}{T + N + \sum_a T_a}.$$

For each $p_m \in \mathbf{P}_Q$:

5. Compute the updated quantile estimate $Q_m$ by inverting $F^{\pm}(x)$ using (3) where the definitions of the bracketing values $x^+$ and $x^-$ are unchanged.

Finally, refill $\mathbf{Q}$ with the updated quantile estimates, clear $\mathbf{D}$, and resume accumulating new records.

When the full set of agent records has been processed, a *server record* is produced to summarize the result. The server record consists of $T$, $N_A$ and a subset of the quantile estimates in $\mathbf{Q}$, including the minimum and maximum values. A set of server records of this form can be aggregated further by applying the IQ server algorithm a second time. Aggregation can thus proceed hierarchically, as Section 8 illustrates.

## 6. ALGORITHM ENHANCEMENTS

Increasing the sizes of $\mathbf{D}$ and $\mathbf{Q}$ improves accuracy. A larger $\mathbf{D}$ allows the subtle features of the

TABLE 1
*IQ estimated quantiles compared to empirical quantiles (EQ) of the 41,928 transaction times illustrated in Figure 2*

| Quantile | 0.5 | 0.75 | 0.9 | 0.95 | 0.99 | 0.995 |
|---|---|---|---|---|---|---|
| IQ | 190 | 323 | 821 | 1338 | 4674 | 5154 |
| EQ | 189 | 320 | 826 | 1280 | 4807 | 5147 |
| Difference | 1 | 3 | −5 | 58 | −133 | 7 |
| $2 \times$ s.e.(EQ) | 1.3 | 5.4 | 32 | 72 | 134 | 130 |

For IQ, $\mathbf{D}$ and $\mathbf{Q}$ both have size 100. Absolute differences between IQ and EQ are less than two standard errors of the empirical quantiles.

underlying distribution to be better represented in the empirical CDF before folding into $\mathbf{Q}$. A larger $\mathbf{Q}$ reduces interpolation errors because interpolation is used over shorter intervals.

If memory cannot be increased, it is sometimes desirable to sacrifice accuracy in the central quantiles for improved accuracy in the tails. This tradeoff can be achieved by manipulating the probability levels $\mathbf{P}_Q$ associated with quantiles in $\mathbf{Q}$. Generally, if good accuracy is desired for a quantile with probability level $p$, then it is helpful for $\mathbf{P}_Q$ to place probability values more densely near $p$. But focusing on $p$ leaves fewer probabilities elsewhere with the result that, while accuracy of the $p$th quantile improves, accuracy of other quantiles degrades. We have used probability levels that are either uniformly spaced between 0 and 1 or uniformly spaced on the scale $\log(p/(1-p))$, as in the example in Section 4.3.

Tail quantile accuracy may also be improved by applying nonlinear interpolation to $\mathbf{Q}$, which is equivalent to applying linear interpolation to a transformation of $\mathbf{Q}$. In most applications it is not feasible to determine an optimal transformation because the shape of the distribution is unknown, so it is often desirable to choose a transformation that performs well over a wide variety of datasets. The performance study in Section 7 compares uniformly spaced probability values and linear interpolation with logit spaced probability values and *logit interpolation*, which is defined by taking

$$\mathtt{interp}(x, x_0, x_1, p_0, p_1)$$
$$= g^{-1}\left(g(p_0) + (g(p_1) - g(p_0))\frac{x - x_0}{x_1 - x_0}\right)$$

in (1) and

$$(4) \qquad \rho = \frac{g(F^+(x^+)) - g(p_m)}{g(F^+(x^+)) - g(F^-(x^-))}$$

in (3), where $g(p) = \log(p/(1-p))$ is the logit function and $g^{-1}(x) = 1/(1 + \exp(x))$ is its inverse. In principle, $g$ should be chosen so that $g(F(x))$ is nearly linear, but $F$ is unknown. Although logit interpolation may not be optimal, it should be better than linear interpolation if exponential tails are expected.

## 7. PERFORMANCE OF THE AGENT ALGORITHM

The core of our network monitoring methodology is the IQ agent algorithm that computes incremental quantiles from raw data. To study its performance, we simulated it with $\mathbf{D}$ and $\mathbf{Q}$ of size 41 each. Linear interpolation with uniform probability values (shown as inner ticks along the top axes in Figure 6) and logit interpolation with logit probability values (inner ticks along the bottom axes) were used in the simulation. The logit probabilities are actually at 41 convenient round values that are approximately uniformly spaced on the logit scale. Three distributions are considered: the standard normal, standard log-normal and beta$(9, 2)$, which has a very long left tail and sharp rise to a mode in the right tail. Quantiles were estimated after 1000 and 10,000 independent observations, which implies that the buffers were emptied 24 and 243 times, respectively, and then one more time at the 1000th and 10,000th observations, respectively.

Simulated performance is measured by the ratio of the root mean squared errors (RMSEs) of the IQ and empirical quantile (EQ) estimates where the RMSEs are computed over 1000 runs of the simulation. The horizontal axes in Figure 6 are on the logit scale to show the behavior of the extreme quantiles.

Not surprisingly, Figure 6 shows that uniformly spaced probability values and linear interpolation perform poorly in the tails of the normal distribution. At $N = 1000$ and $p = 0.005$, the IQ RMSE is about four times the EQ RMSE. Moreover, relative performance degrades with $N$. By $N = 10{,}000$ the IQ RMSE is about 20 times larger than the EQ RMSE. Plots not shown here suggest that this degradation is due to the bias in the IQ estimates which does not diminish with $N$. Similarly, Figure 6 shows that linear interpolation and uniformly spaced probability values do not provide good performance in the long right tail of the log-normal and the long left tail of the beta, and that performance relative to the EQ estimates degrades with $N$, again due to bias. At the 0.99 quantile of the log-normal, the ratio of RMSEs is about 15 for $N = 1000$ and about 75 for $N = 10{,}000$. A similar pattern is seen near the 0.01 quantile for the beta distribution. That is, when the uniform scale does not tame the tails of a distribution sufficiently, the IQ estimates with uniform probabilities and linear interpolation may be noticeably worse than the empirical quantiles. The RMSEs of the IQ and EQ estimates are nearly identical for the most extreme quantiles under all distributions because these are computed from the minimum and maximum data values, which the IQ algorithm keeps in $\mathbf{Q}$.

For logit probability values and logit interpolation, there are ripples in the ratio of IQ RMSE

to EQ RMSE in the center of the log-normal and beta distributions. These ripples become more pronounced with increasing $N$. The low points of the ripples occur for quantiles that are kept in $\mathbf{Q}$, while the high points are between adjacent quantiles. Degrading relative performance with increasing $N$ is again due to the bias in the IQ estimates that occurs in regions where the density changes rapidly with respect to the logit-spaced probability levels. But in all cases, the IQ RMSE is within a factor of 2 of the EQ RMSE even though the IQ algorithm never computes with more than 82 data values while the empirical quantiles require knowing all 1000 or 10,000 data values at once. In this sense, the IQ algorithm produces usable estimates over a range of distributions.

A second simulation experiment with log-normal data, logit-spaced $p$'s, logit interpolation, and $\mathbf{D}$- and $\mathbf{Q}$-buffers of size 1000 was run to focus on the behavior of IQ estimated quantiles for large $N$. The quality of the IQ estimates was evaluated at $N = 10^K$, for $k = 3, 4, \ldots, 7$. For all values of $k$, the IQ RMSE tracked the EQ RMSE closely in the middle of the distribution. For instance, the ratio of IQ RMSE to EQ RMSE averaged over the middle 95% of the log-normal, $p \in (0.025, 0.975)$, increases from 1.00000 at $N = 10^3$ to 1.01338 at $N = 10^7$, an increase of only about 1%. The ratio of IQ RMSE to EQ RMSE does increase more with $N$ in the tails. For example, at $p = 0.99$ the ratio increases 31.5% as $N$ increases from $10^6$ to $10^7$, but even this bias would not make the IQ estimates unusable in our application. Thus, the IQ estimates are adequate if the probability levels for the $\mathbf{Q}$ and interpolation schemes are suitable.

## 8. PERFORMANCE OF THE AGGREGATED GROUP QUANTILES

Networked software monitoring focuses on the quantiles of the performance experienced by groups of users. We explore the behavior of the aggregated quantiles that are computed by the IQ server algorithm in this section.

*Transaction Time Data.* The data shown in Figure 2 represent 41,928 email transactions for 15 corporate users over one month. Hourly sets of quantiles were computed for each user, and the hourly user quantiles were aggregated to produce hourly records for the group of 15 users. Finally, the hourly group records were aggregated to produce daily quantile estimates for the group.

The IQ agent (user) and server (group) algorithms both used $\mathbf{D}$- and $\mathbf{Q}$-buffers of size 100 with uniformly spaced probabilities $\mathbf{P}_Q$ and linear interpolation on log transaction times. Each agent and server record contained only 11 quantiles corresponding to probability levels $\mathbf{P}_R = \{0, 0.05, 0.10, 0.25, 0.50, 0.75, 0.90, 0.95, 0.99, 0.999, 1\}$.

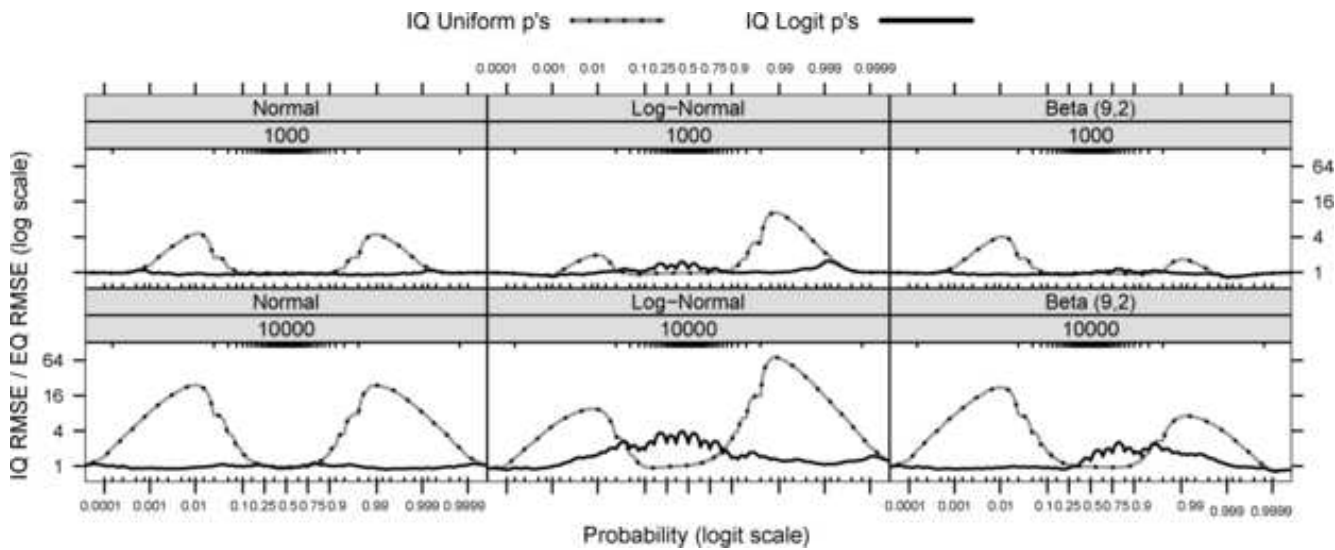

FIG. 6. *Performance of IQ on three distributions. Logit p's combined with logit interpolation perform well in the tails but generally not as well as uniform p's and linear interpolation for the center of the distribution. As the sample size increases from 1000 to 10,000, IQ performance degrades relative to empirical quantiles because the IQ estimates are biased whereas empirical quantiles are not.*

Figure 7 compares the incremental and empirical quantiles for a week of hourly group records and Figure 8 compares these quantiles for a month of daily aggregates. In each figure, the lower line tracks empirical medians and the upper line tracks empirical 0.9 quantiles. Darker vertical segments connect empirical quantiles to the corresponding IQ estimates, so longer lines correspond to poorer IQ estimates. Figure 7 represents results after two stages of processing: one at the agent and one at the server. Figure 8 shows results after an additional application of the IQ server algorithm to compute daily quantiles.

The IQ estimates track empirical quantiles reasonably well, especially at the daily level where most differences are imperceptible. At the hourly level, some errors in the 0.9 quantiles are noticeable, but this reflects the limits on the accuracy that can be achieved when each agent record consists of only 11 quantiles. Table 2 reports the fraction of cases in which incremental quantiles were within 10% of the correct empirical values.

*Simulated Inhomogeneous Agents.* The data from different users of networked applications are typically not homogeneous because their network environments and software usage differ. Here we report the results of a simulation that gives some insight into how the IQ algorithm responds to outlying users. These results also address the question of whether the order in which the records from inhomogeneous agents are received matters, given that the server processes records sequentially. The following simulation is meant to be realistic, but only exemplary because it is not possible to test or even specify the full range of conditions that could be encountered in a real network monitoring application.

In the simulation, agent records of length $I = 10$ (i.e., 10 quantile estimates, not raw data values) are constructed for 1000 agents independently: 99% of the agents are *nominal* and 1% are *outlying*. In either case, the simulated record $\mathbf{R}_a$ for agent $a$ ($a = 1, \ldots, 1000$) is formed as follows. First an i.i.d. sample of $T_a = 1000$ values is drawn from a log-normal (base 10) distribution:

$$X_{a,t}|m_a \sim 10^{N(m_a, V_1)}, \quad t = 1, \ldots, 1000.$$

The agent record consists of $I = 10$ empirical quantiles $\mathbf{R}_a = (R_{a,1} \leq \cdots \leq R_{a,10})$ corresponding to probabilities of 0, 1 and eight values equally spaced between 0.005 and 0.995 on the logit scale. The medians $m_a$ of the logged agent distributions are independent and log-normally distributed:

$$m_a|M_a \sim 10^{N(M_a, V_2)},$$

where

$$M_a = \begin{cases} 0, & \text{with probability 0.99}, \\ 2, & \text{otherwise.} \end{cases}$$

Nominal agents are those with $M_a = 0$; outliers are those with $M_a = 2$. We set $V_1 = V_2 = 0.0924$, resulting in

$$\frac{Q(0.99|M_a)}{Q(0.01|M_a)} = 100 \quad \text{for } M_a = 0 \text{ and } 2,$$

where $Q(p|M_a)$ is the $p$th quantile of $[X_{a,t}|M_a]$. That is, the central 98% of nominal data cover two orders of magnitude, as do the central 98% of outlying data. Furthermore, with $M_a$ taking values of 0 and 2, the outlying data are centered two orders of magnitude larger than the nominal data. The complete mixture covers about four orders of magnitude between its 0.01 and 0.999 quantiles. Note, however, that agents are not homogeneous. Both nominal and outlying agents have random medians and thus each agent record summarizes a different distribution of data. Agent records constructed using empirical quantiles as above do not have any errors associated with agent-level IQ estimation. Thus, this simulation only considers performance of the server-level algorithm.

At the server, the **D**-buffer is sized to hold 100 length-10 records and the **Q**-buffer holds 1000 quantile estimates with probabilities of 0, 1 and 998 values equally spaced between $10^{-6}$ and $1 - 10^{-6}$ on the logit scale. Interpolation uses $g(\cdot) = \text{logit}(\cdot)$ as described in Section 6.

Figure 9 plots the ratio of IQ RMSE to EQ RMSE after processing the agent records representing, in aggregate, 1000 data values for 1000 agents, or one million data values in all. The plot has two curves, one for aggregation on the nominal data scale (solid line) and one for aggregation of logged agent records (dotted line). Logit interpolation is used in both cases. The most obvious feature is that transforming the data to the log scale improves performance, especially in the central part of the distribution. In fact, the worst relative performance occurs near the median when aggregating nominal data, but with logged data the IQ median estimate has the same RMSE as the empirical median.

Both curves in Figure 9 show that the far upper tail, corresponding to the 1% of outlying agents, is estimated with essentially the same accuracy as empirical quantiles. This is not a trivial result because, even with logged data, each agent describes a different distribution and the complete mixture is *not* Gaussian. Some additional experimentation showed that nominal-scale performance in the central portion of the distribution can be improved by increasing the agent record length above 10. We chose length-10 records, however, because this closely matched the stringent requirements imposed for monitoring networked applications.

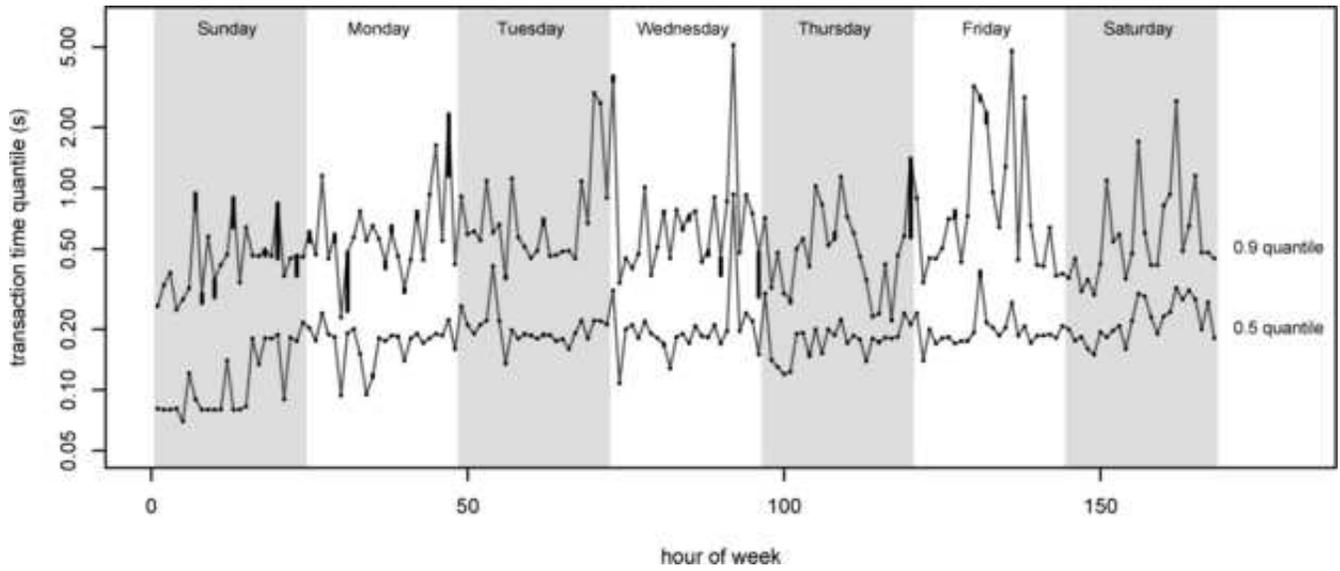

Fig. 7. *Hourly quantiles of email transaction times over a one-week period. Lines track empirical 0.5 and 0.9 quantiles while vertical bars connect empirical quantiles to IQ estimates in order to highlight differences. Two rounds of IQ were performed: first, agents prepared hourly records; then the server combined agent records to obtain the aggregate hourly results shown.*

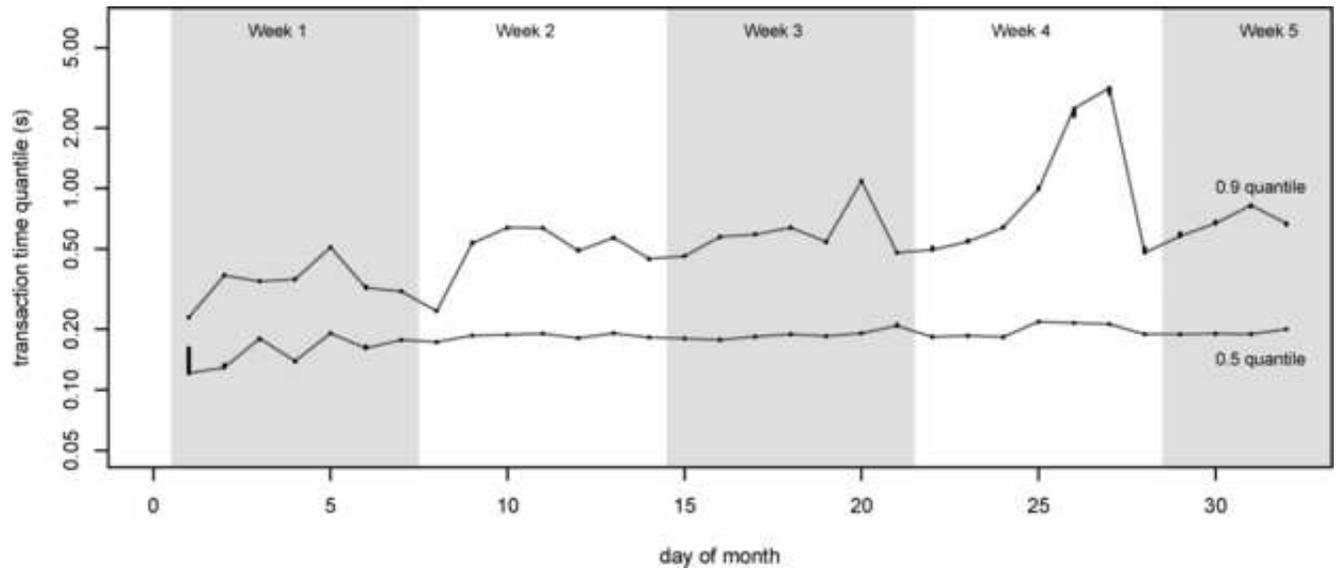

Fig. 8. *Daily 0.5 and 0.9 quantiles of email transaction times over a one-month period. IQ results are obtained from aggregating hourly records such as displayed in Figure 7, which corresponds to Week 3 in this figure.*

Table 2
*Fraction of cases in which IQ estimates are within 10% of the empirical quantiles for email transaction times*

| Aggregation level | Number of cases | Fraction within 10% | |
|---|---|---|---|
| | | 0.5 quantile | 0.9 quantile |
| Hourly | 768 | 0.999 | 0.929 |
| Daily | 32 | 0.969 | 1.000 |

As a second experiment, we fed the agent records to the server algorithm sorted by increasing values of their log-medians $m_a$ rather than in random order. In particular, most outlier records were processed *after* nearly all nominal records had been processed. Remarkably, performance curves (not shown) for aggregating the ordered records are indistinguishable from the curves of Figure 9. In this experiment, at least, it made no difference whether inhomogeneous agent records were presented in random or sorted order.

## 9. DISCUSSION

Most corporate software is highly reliable, so it is only the tail behavior (and, hence, tail quantiles) of performance data that are of interest. Moreover, software performance and reliability are often monitored for groups of users, not individual users, partially because any one user may access the software so infrequently that statistics based on individual users are too unreliable to be interesting. Thus, monitoring the reliability and performance of networked applications naturally leads to distributed monitoring and aggregating quantiles over groups of users and time. We have presented one approach to estimating aggregated quantiles from distributed monitoring data, and shown that it can give trustworthy estimates using limited agent and network resources even if the agents are not homogeneous and their records arrive in what seems to be perverse (smallest first) order.

While this paper has focused on networked software, the need for estimating aggregated quantiles for highly reliable business systems arises in other contexts, too. Examples include communications software that routes calls to appropriate support staff in technical help centers and package tracking software used by delivery services to route shipments at waypoints in a network of transit sites. Each of these applications can generate huge amounts of data such as transaction time, size and completion status that can be used to monitor performance and reliability. For example, the call center for one computer manufacturer has on the order of 10,000 agents that together handle millions of transactions per day, each of which can, in principle at least, be monitored for setup and response time. The transactions for an agent can be measured, quantile records computed, and then aggregate performance by work group or location can be estimated.

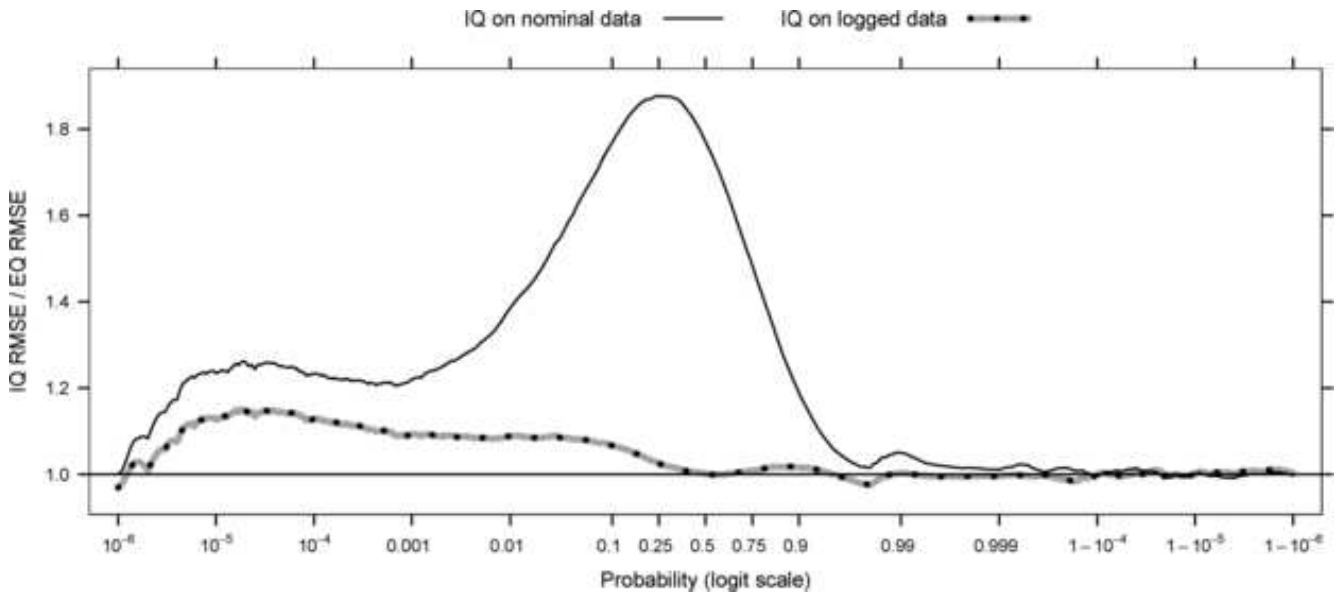

FIG. 9. *Server performance on inhomogeneous agents. The server processes* 1000 *length-*10 *agent records, each of which summarizes* 1000 *data values. Marginally, a data value from the group of agents follows a mixture of log-normal distributions that covers four orders of magnitude. Agent records are processed at the server in batches of* 100 *using a* **Q**-*buffer of length* 1000. *The resulting RMSEs are less than twice those of empirical quantiles for nominal-scale updating and less than* 120% *of the EQ RMSEs with log-scale updating. Results are averaged across* 500 *simulation runs.*

This paper has shown that IQ estimation provides a way to track performance at several levels of aggregation over time, agents or space simultaneously, where the set of agents, portion of the network, and time period of interest are not necessarily fixed in advance. Although IQ estimation can be applied whenever multiple quantiles are needed, it is probably most useful when interest focuses on tail quantiles or the data are not expected to follow a parametric distribution. This paper shows that IQ estimates provide useful information throughout the range of the data if logit probability values are combined with logit interpolation. This is especially important for evaluating the reliability and performance of networks and other systems that nearly always perform well. For such systems, only tail quantiles are of interest.

The IQ method can be characterized as "quick and dirty" in the sense that we work under the tight computational constraints imposed by the application, notably the fixed sizes of buffers and summary records and the desire for simplicity. We are also willing to proceed with a method whose conventional statistical properties (e.g., bias and convergence) are not yet fully understood, partially because standard sampling and distributional assumptions seem unlikely to hold in the motivating applications. As would be expected with a quick and dirty method, there are limitations to the resulting estimates. For example, they assume that interest centers on aggregate performance over the entire workgroup or reporting interval rather than on the details of the performance experienced by individual users during the interval. Similarly, IQ estimates do not take account of trends over time or time-of-day patterns, such as the difference between peak and off-peak hours. It would be straightforward to allow trends by incorporating exponential weighting into the averaging steps for updating $\mathbf{Q}$. Time-of-day or day-of-week patterns could be incorporated by starting each reporting period with a $\mathbf{Q}$ specific to the time period instead of an empty buffer or one that is continuously updated over all time periods. These can also be accommodated by defining the duration over which a $\mathbf{Q}$-buffer is filled. Longer periods give more stable estimates, but may include data with dissimilar distributions.

On a mixed distribution with spikes, some empirical quantiles will be exactly correct with high probability in large samples. IQ estimates do not behave as well, but if the spikes are known in advance, then the IQ algorithm could be easily modified to count hits at the spikes separately and process the remaining data through the IQ algorithm. There are, for example, preferred packet sizes in network data that cause spikes in the size distributions, but these are known in advance and so can be planned for.

Finally, the spacing in the probability values affects the performance of IQ estimates, but our algorithm makes no attempt to adjust the probability values over time. An algorithm that adjusted the probability values to minimize interpolation error associated with $F_Q$ would perform better, but probably not be as quick or straightforward. A simpler approach would be to collect some training data to get a ballpark estimate of the shape of the distributions of interest and use that shape to inform the choice of probability values for $\mathbf{Q}$. If extreme tails are of interest, it may help to gradually extend the most extreme probabilities into the tails as the total sample size builds. For example, the smallest nonzero probability could be maintained at approximately $0.5/T$, and nearby probabilities could be adjusted correspondingly.

While there are many ways in which IQ estimates could be improved, the fact that they are easy to explain, easy to interpret, easy to implement, and provide useful information about tail behavior, even for aggregates over time, users and space, makes IQ estimates an attractive choice for monitoring performance and reliability.

## ACKNOWLEDGMENT

The problem of monitoring networked applications was first brought to our attention by engineers in a business unit of Lucent Technologies when we were all in the Statistics Research Department of Bell Labs. The approach developed in this paper was designed to meet their needs and has been implemented by them.